\documentstyle[aps]{revtex}
\input psfig.sty
\newcommand{\be}{\begin{eqnarray}}
\newcommand{\ee}{\end{eqnarray}}
\newcommand{\ben}{\begin{eqnarray*}}
\newcommand{\een}{\end{eqnarray*}}
\newcommand{\bc}{\begin{center}}
\newcommand{\ec}{\end{center}}
\newcommand{\Vo}{(4V_{o}^{3}/3)^{1/3}}

\newcommand{\as}{\alpha_s}
\newcommand{\sinfi}[1]{\sin\varphi({#1})}
\newcommand{\cosfi}[1]{\cos\varphi({#1})}
\newcommand{\ups}{\upsilon(p)}
\newcommand{\ds}{\displaystyle}

\newcommand{\Vpq}{V(|p-q|)}

\def\thebibliography#1{\section*{References}\list
 {\arabic{enumi}.}{\settowidth\labelwidth{[#1]}\leftmargin\labelwidth
 \advance\leftmargin\labelsep
 \usecounter{enumi}}
 \def\newblock{\hskip .11em plus .33em minus -.07em}
 \sloppy
 \sfcode`\.=1000\relax}

\begin{document}
\title{\bf  The quark inside hadron.}
\author{{\bf Obid M. Juraev} \\
{\normalsize International Centre for Theoretical Physics, Trieste 34100,}
{\bf Italy}\\
{\normalsize and}\\
\normalsize Institute of Nuclear Physics, Tashkent 702132,
{\bf Uzbekistan}\thanks{Permanent Address} }

\date{07 September  1998}
\newpage

\maketitle
\begin{abstract}
We study and solved Schwinger-Dayson equation for massive quarks in the
quark model. Interaction potential is choose as sum oscillator and
Coulomb terms. The calculation show us that negative energy of quark exist
under $m_{o} \ll \Vo$ i.e. when mass quark is less that parameters of
scale. Us are offered to use asymptotical behaviour at the solutions
Schwinger-Dayson equation selection physical and non-physical conditions.
The quark condensate under different values mass quarks are calculated.
The scale parameters $\Vo$= 520 MeV. 
\end{abstract}

\newpage
\section{Introduction.}

For description of  decays of particles to us needed will know a behaviour
of wave functions and the spectra of masses of mesons and baryons. In
literature there are several general investigation ways:
quark model, lattice QCD, bag model, QCD sum rules, quenched QCD and
other phenomenological models and approaches. They have their own advantages and 
defect.

The most real approach is quark model based on Schwinger-Dyson (SD)
and Bethe-Salpeter (BS) equations. This quark model proposed by the Orsay group \cite{Orsay},
perfection by Pervushin
\cite{Perv} and Bicudo-Ribeiro \cite{Big}. The main purpose the works
\cite{Orsay,Perv,Big} 
was offer a method define wave functions and obtain spectrum masses of mesons 
and hadrons. Next step development of quark model is calculate meson
decays \cite{Jur}  and improve upon \cite{Rob}. Recently in the paper
\cite{Szcz} presented the  development of the quark model
 and they have
point that model a pretender on the description hadron physics.

Thereby we have quark potential model which based on SD and BS equations.
The definition 
of the form of quark-antiquark potential is one of the actual problem of
the meson spectroscopy.

In the present paper we investigate quark mass function based on SD equation. We 
consider quark in hadron with interaction potential chosen as sum Coulomb and oscillator
terms. We has solve SD mass gap equation and  the energy of quarks are calculated in this
approach. Using asymptotic behavior when $p\rightarrow\infty$ we can select solution on
physical and non-physical stages of quarks. By means of solutions SD equation we have
calculate $<{\bar q} q>$ quark condensate under different masses of
quarks.

Let us consider the effective covariant relativistic action for singlet
channel  \cite{Perv}:

%                  Formula  2
\be
S_{eff}=\int d^{4}x \bar q(x) (i \rlap/\partial - \widehat m_{o}) q(x) -
\frac{1}{2} \int d^{4}x d^{4}y
q_{\beta_{2}}(y)\bar q_{\alpha_{1}}(x) \nonumber\\
\times \left[ K_{1}(x-y|\frac{x+y}{2})
\right]_{\alpha_{1},\beta_{1},\alpha_{2},\beta_{2}}
q_{\beta_{1}}(x)\bar q_{\alpha_{2}}(y)
\label{eq:formla2}
\ee
with interaction (kernel) \\
\bc
$K_{1}(x-y|X)_{\alpha_{1},\beta_{1}/
\alpha_{2}\beta_{2}}= \rlap/\eta_{\alpha_{1}\beta_{1}}
[V(z_{\perp}\delta(z\cdot\eta)]
{\rlap/\eta}_{\alpha_{2}\beta_{2}} $
\ec
where $\widehat m_{o}=(m_{o1},...,m_{o{n_{f}}})$ are the bare masses,
$\alpha$ or $\beta$ are the short notation for the Dirac and flavour
index.
Quantization axis is chosen 
\be 
\eta_{\mu} \sim \frac{\ds 1}{\ds i} \frac{\ds \partial}{\ds\partial X_{\mu}},
\label{Mark:Yukava}
\ee
where $\eta_{\mu}$ -- single time-axis.
 From the covariant effective action (\ref{eq:formla2}) we can obtain SD equation
for  mass operator ($\Sigma(p^{\perp})$) depends on transversal momentum
$p^{\perp}$ with  nonzero mass current quark  ($m_o$) \cite{Perv}.
\be
\Sigma(p_{\perp}) = m_o + i \int \frac{d^4 q}{(2\pi)^4}V({\vec 
p}_{\perp}-{\vec
q}_{\perp}){\rlap/ \eta} G_{\Sigma}(q) {\rlap/\eta}
\label{green1}
\ee
where $\rlap/\eta = {\eta}^{\mu}\gamma_{\mu}$, 
 ${\eta}_{\mu}$
% - time- axis  (${\eta}^{2}=1$),
 is satisfy Markov-Yukava principle
(\ref{Mark:Yukava}), \\
${\vec p}^{\perp}_{\mu} = {\vec p}_{\mu} - {\eta}_{\mu}({\eta} {\cdot}
{\vec p})$- momentum, which orthogonal on time axes (${\vec p}_{\perp}, \eta=0$), \\
 and $G_{\Sigma}$ -  Green's function of quark
\be
G_{\Sigma}(q) = [\rlap/ q - \Sigma(q)]^{-1}
\label{green2}
\ee
%The equation is diagonal in flavour space, and then wenotwriteflavourindexes.

 Let us note that after integration (\ref{green1}) over "time" component 
($q^{\mu}\equiv q
\cdot \eta$) we obtained the mass operator with  perpendicular
component only.
We can write in Lorentz covariant presentation :
\be
\Sigma({\vec p}_{\perp}) = {\rlap/{\vec p}}_{\perp} + E({\vec p}_{\perp})
S^{-2}({\vec p}_{\perp})
\label{green3}
\ee
where $E$ -  energy of quark, and $S$ - unitary operator which can to
present(parametrize) through $\upsilon$ function as
\be
S({\vec p}_{\perp}) = \exp\{-{\rlap/{\vec p}}_{\perp}\upsilon({\vec
p}_{\perp})\}
\label{green4}
\ee
where ${\vec p}_{\perp}={{\vec p}_{\perp}}/{ | {\vec p}_{\perp} | }$ ,
$ | {\vec p}_{\perp} | = \sqrt{({\vec p}_{\perp})^2}$. 
The one-particle Green function (\ref{green2}) can be represented in the
form of the expansion over the states with positive and negative energies
\be
G_{\Sigma}(q)=\left[\frac{{\Lambda}_{(+)}^{\eta}(q_{\perp})}
{q^{\eta}-E(q_{\perp})+ i\epsilon} +
\frac{{\Lambda}_{(-)}^{\eta}(q_{\perp})}{q^{\eta}+E(q_{\perp})+
i\epsilon}\right]{\rlap/
\eta}
\label{green5}
\ee
\be
\Lambda_{(+)}^{\eta}=S(q^{+})\Lambda_{(+)}^{\eta}(0)S^{-1}(q_{\perp})
\label{green6a}
\ee
\be
\Lambda_{(-)}^{\eta}=S(q^{+})\Lambda_{(-)}^{\eta}(0)S^{-1}(q_{\perp})
\label{green6b}
\ee
 where $\Lambda_{(\pm)}^{\eta}(0)$ are  projection operators of
positive
and
negative
value,
\be
\Lambda_{(+)}^{\eta}(0) = \frac{1}{2}({\bf I} + {\rlap/ \eta}),
\label{green7a}
\ee
\be
\Lambda_{(-)}^{\eta}(0) = \frac{1}{2}({\bf I} - {\rlap/ \eta}),
\label{green7b}
\ee
here $\bf I$ --  one-dimensional 4-matrix.

Since energy of quark $E(0) \sim m_o$, $E(\infty) \sim p_{\perp}$
% and function $\upsilon(0)=\frac{\pi}{2}$, $\upsilon(\infty)=0$
 and considering  that $S^2(0)=1$, $S^2(\infty)=-1$, so propagator of quark in limit
$p_{\perp}\rightarrow 0$ coincide with propagator free quark in rest  system
\ben
G(0)=-\frac{1}{2}\left[\frac{1+{\gamma}^{o}}{p_{||}-m-i\epsilon}-
\frac{1-{\gamma}^{o}}{p_{||}+m-i\epsilon}\right]
\een
In limit $p_{\perp}\rightarrow\infty$ propagator $G(\infty)$ 
coincide with propagator massless quark
\ben
G(\infty)=\frac{1}{2}\left(
\frac{1+{\vec\gamma}{\hat{\vec p}}}{p_{||}-p_{\perp}-i\epsilon}-
\frac{1-{\vec\gamma}{\hat{\vec p}}}{p_{||}+p_{\perp}-i\epsilon}
\right)
\een

Let us  consider SD equation (\ref{green1}) in rest system ${\eta}_{\mu}=(1,0,0,0)$,
of bound state, one of the praticles is quark. Inserting 
(\ref{green3})-(\ref{green7b}) into
(\ref{green1}) and after integration over  $q^{\eta}=q^{0}$ we have get  next equation
for $\upsilon$ function
\be
m_{o}\sin2\ups - p\cos2\ups=\frac{1}{2}\int\frac{d^3p}{(2\pi)^3}V(|{\vec 
p}-{\vec
q}|)[\cos2\upsilon(q)\sin2\ups -   \nonumber\\
\frac{{\vec p}\cdot{\vec
q}}{pq}\sin2\upsilon(q)\cos2\ups ]
\label{green8}
\ee
and expression for  energy of quark 
\be
E(p)=m_{o}\cos2\ups+ p\sin2\ups +
\frac{1}{2}\int\frac{d^3p}{(2\pi)^3}V(|{\vec
p}-{\vec q}|)[cos2\upsilon(q)\cos2\ups- \nonumber\\
\frac{{\vec p}\cdot{\vec
q}}{pq}\sin2\upsilon(q)\sin2\ups]
\label{green9}
 \ee

\section{Schwinger-Dyson mass gap  equation.}
In the literature the equations (\ref{green8}),(\ref{green9})  usually called
as mass gap equation. Lets us make 
$\ups=\frac{1}{2}[\frac{\pi}{2}-\varphi(p)]$ variable change in
(\ref{green8})(\ref{green9}). So we obtain the following equations
%                  Formula  3
\be
\begin{array}{lcl}
E(p)\sinfi{p} & = &m_{o}+
\ds{\frac{1}{2}} \ds{\int}\frac{\ds d^3q}{\ds (2\pi)^3}
 \Vpq \sinfi{q} \\
E(p)\cosfi{p}& = &p+\ds{\frac{1}{2}} \ds{\int} 
\frac{\ds d^3q}{\ds (2\pi)^3}
\Vpq \hat p \hat q \cosfi{q}
\end{array}
\label{eq:formla3}
\ee
where 
%$(dq)=\frac{\ds d^3q}{\ds (2\pi)^3}$,
 $p=|\vec p|$, 
${\hat p} =\frac{\ds \vec p}{\ds p}$.

The SD equation define one-particle energy of flavour quark.
The potential phenomenology of spectroscopy of quarkonia \cite{Jur} in
the meaning of the real interaction, uses in general, the sum Coulomb
and the increasing potential. In the increasing potential is used the lattice 
calculation for description of the heavy quarkonia.
The definition of the form ${\bar q} q$ -- potential is one of the actual
problem of the meson spectroscopy. We chose the potential
in the form as sum of oscillator and Coulomb terms.
%                  Formula  4
\be
V(p)= - \frac{4}{3} \left[\frac{4\pi \alpha_s}{p^2} +
(2\pi)^3 V^{3}_{o}\triangle_{p}\delta^3(p)\right]
\label{potential}
\ee

One particle SD equation (\ref{eq:formla3}) for quark in dimensionless
units takes a form
%                  Formula  5
\be
\frac{d^2\varphi(p)}{dp^2}+\frac{2}{p} \frac{d\varphi(p)}{dp} +
\frac{\sin 2\varphi(p)}{p^2}-2p\sinfi{p}+2m_{o}\cosfi{p} +\nonumber\\
+\frac{2\as}{3\pi}\int dq \frac{q}{p}
\left\{{\cal R}_1(p,q)
%\left[\sinfi{q}-\frac{m_{o}}{r(q)}\right]
\cosfi{p} -
{\cal R}_2(p,q)
%\left[\cosfi{q}-\frac{q}{r(q)}\right]
\sinfi{p}\right\} = 0
\label{eq:formla5}
\ee
%                  Formula  6
\be
E(p)&=&p\cosfi{p}+m_{o}\sinfi{p}-\frac{1}{2}[\varphi'(p)]^2- 
\frac{\cos^2\varphi(p)}{p^2}\nonumber\\
& &+\frac{2\alpha_s}{3\pi} \int dq \frac{q}{p}
\left\{{\cal R}_1(p,q)
%\left[\sinfi{q}-\frac{m_{o}}{r(q)}\right]
\sinfi{p} -
{\cal R}_2(p,q)
%\left[\cosfi{q}-\frac{q}{r(q)}\right]
\cosfi{p}\right\}
\label{eq:formla6}
\ee
where
\ben
{\cal R}_1(p,q)&=&
\ln{\left| {p+q \over p-q} \right|}
\left[\sinfi{q}-\frac{m_{o}}{r(q)}\right], \\
{\cal R}_2(p,q)&=& \left({p^2 +q^2 \over 2pq}
\ln{\left| {p+q \over p-q} \right|}-1\right)
\left[\cosfi{q}-\frac{q}{r(q)}\right],
\een
$r(q)=\sqrt{q^2+(m_{o})^2} $.
The equation (\ref{eq:formla5}) and expression for self-energy of 
quarks (\ref{eq:formla6}) is present in dimensionless unit $\Vo=1.$ We can  define this
parameter
in physical region after solve BS equation, from spectroscopy 
mesons and baryons. In literature \cite{Orsay},\cite{Big},\cite{Jur} usually
use
as:
$\Vo\approx 280 \div 480$
MeV.
\section{Numerical method.}
To solve the equation (\ref{eq:formla5}) by analytical method is difficult
that is way we'll
find the solution it by use numerical method. Rewrite equation
(\ref{eq:formla5}) in following type:
%
%                 Formula 5o
%
\be
{\cal F}_1[\varphi(p)]+{\cal F}_2[\varphi(p)]=0 ,  
\label{eq:formla5o}
\ee
where $${\cal F}_1[\varphi(p)]=
\frac{d^2\varphi(p)}{dp^2}+\frac{2}{p} \frac{d\varphi(p)}{dp} +
\frac{\sin 2\varphi(p)}{p^2}-2p\sinfi{p}+2m_{o}\cosfi{p},$$
$${\cal F}_2[\varphi(p)]=
\frac{2\as}{3\pi}\int dq \frac{q}{p}
\left\{{\cal R}_1(p,q)
%\left[\sinfi{q}-\frac{m_{o}}{r(q)}\right]
\cosfi{p} -
{\cal R}_2(p,q)
%\left[\cosfi{q}-\frac{q}{r(q)}\right]
\sinfi{p}\right\}.$$
The problem can be solved step by step scheme:
{\begin{description}
\item[{\it i)}] $\alpha_s=0, m_o=0, \Vo=1$. The current quark mass equal null end
equation (\ref{eq:formla5o}) disregarding  Coulomb interaction potential
is considered.
 We have chosen interaction potential as increasing oscillator potential.
\item[{\it ii)}] $\alpha_s=0, m_o\neq0, \Vo=1$. The equation
(\ref{eq:formla5o}) with provision for current quark mass on the intervale
$0\leq m_o \leq M_{\it Large mass quark}$.
\item[{\it iii)}] $\alpha_s\neq 0, m_o \neq 0, \Vo=1$ The equation
(\ref{eq:formla5o}) oscillator plus Coulumb term of 
interaction potential and current quark mass not equal zero.
\end{description}}
Unfortunately to consider the equation (\ref{eq:formla5o}) for case
{\it iii)} without oscillator terms of interaction potential (i.e.
$\Vo=0$) to us it was not possible. 

We have derived numerical solutions to the integr-differential equation
(\ref{eq:formla5o})
with boundary conditions,
\be
\varphi(0)={\pi\over 2}\ \ \ ,\ \ \ \ \  \varphi(\infty)=0
\label{eq:formla7}
\ee
using the computational scheme developed in ref.\cite{Newt}. This scheme
consist the 
modification
of the Newton iterations by combining the continuation over a parameter 
($m_o$ or $\alpha_s$) 
method. This scheme has a convenient algorithm for assignment 
of the initial conditions as functions of the external parameters 
($m_o$ or $\alpha_s$),due to a special choice of the iteration parameter as
a step of the modified Newton iteration. So, in this scheme the solutions to
problem  (\ref{eq:formla5o}),(\ref{eq:formla7})  are used as an input.

	The numerical solutions to boundary problem
(\ref{eq:formla5o}),(\ref{eq:formla7}) for
several values of $m_o$ are shown in Figure 1, under $\alpha_s=0.2$. The values
of free parameters $m_o, \Vo$ and $\alpha_s$ belonging to the physical region 
will be defined from the solutions to the BS equation for mesons.

\section{Results and conclusion.}
We solve SD equation by using numerical method with choose potential (\ref{potential}) for
massive quarks. In the event of massless quarks ($m_o=0$) and without Coulomb 
term of potential our results coincide with \cite{Orsay}. The SD equation have solution 
with some nodes. The asymptotical behaviour $\varphi$ when $p\rightarrow\infty$ is positive
value because $\varphi(p\rightarrow\infty)>0$
for massles quark and $\varphi(p\rightarrow\infty)=\frac{m_{o}}{p}$.
 The authors \cite{Orsay} study asymptotic behaviour when
$p\rightarrow\infty$ for massles quarks and without Coulomb term of
interaction potential.

We consider without nodes solution of SD equation, though needed to note that equation
(\ref{eq:formla5o}) has an infinite number of solutions. From asymptotic behavior $\varphi$ we
are to select on physical and non-physical solutions. Behavior $\varphi$ under 
$p\rightarrow\infty$ following: 
%
%                    Formula 8
%
\be
\varphi(p\rightarrow\infty) \rightarrow \frac{m_o}{p}
\label{eq:formla8}
\ee
and it has positive sign, this fact indicates us that the nodes solution of 
number $n = 0, 2 , 4, ...$ (i.e. even number) they are physical. The nodes solution of 
number $n = 1, 3, 5, ...$ makes no physical sense because asymptotic condition
(\ref{eq:formla8}) is violated.

     We have calculate self-energy of quarks under different parameters
mass, see Figure 2. Studies 
have reveal us behavior $E(p)$ under $m_o > \Vo$ with certain accuracy comply 
with nonrelativistic approximation $\sqrt{p^2+m_o^2}$, authors \cite{Big} were
calculated such results. Presence of negative energy (when $m_o = 0$)
note on \cite{Orsay}.Authors \cite{Big} researched a negative energy and adding in 
the potential of a constant energy shift they verified BS equation for mesons and baryons. In
fact \cite{Orsay} adding to the potential in space coordinates a constant energy shift then the 
quark energy is shifted by exactly half of that constant opposite sign.

	We consider a region when $m_o < \Vo$. On the  figure 2 we see
that negative energy has an end value (under $p=0, m_o=0$) $E(0)=-6\Vo$,
reference to \cite{Orsay}. With increasing masses an energy of quark grows
in the negative area and for $m_o\geq \Vo$ continues to grow in the
positive area.  On the Figure 2 (line (a)) have plot a curved
line ${\bar E}(p)=p-\frac{1}{p^2}$ in the neighborhood $p=0$.

     In our approximation the quark condensate define through the part
nonperturbative Green's function \cite{Perv}
%
%                 Formula 9
%
\be
<{\bar q} q > = i 2 N_c tr \int \frac{d^3
p}{(2\pi)^3}[G_{\Sigma}(p)-G_{m}(p)]= \nonumber\\
= - 2 N_c \int \frac{d^3
p}{(2\pi)^3}[\sin\varphi(p^{\perp}) - \frac{m_o}{\sqrt{p^2 + m_o^2}}] ,
\label{eq:formla9}
\ee
where $G_{\Sigma}(p)=[\rlap/ p - \Sigma(p)]^{-1}$
and $G_{m}(p)=(\rlap/ p - m)^{-1}$ are Green's functions of "dressed"
and
"bare"
quarks respectively, and $\varphi(p)$ - the solution of SD equation of
quark $q$.

In the paper \cite{Orsay}, as we already mentioned earlier, SD
equation is considered in the case $m_o=0$ and they calculated $<{\bar
q}q>= - (140 Mev)^3$ in units $\Vo= 289 Mev$, we have repeat this result.
Moreover have calculate behaviour $<{\bar q}q>$ under different values of
masses $m_o$ and unit parameter $\Vo$, see Figure 3. In the event $\Vo=
520 $ Mev (line (c) in the Figure 3) our results compare excellent with
commonly value of quark condensate (-250 Mev)$^3$ \cite{Shif}. Completely
proper value of scale $\Vo$ will be calculate after solving the equation
BS in our approach.

\section{Acknowledgments}
Author is grateful for hospitality of the High Energy Physics at Abdus
Salam International Centre for Theoretical Physics during a visit in which
this work was conducted.

%\newpage

\newpage
\begin{figure}
\psfig{figure=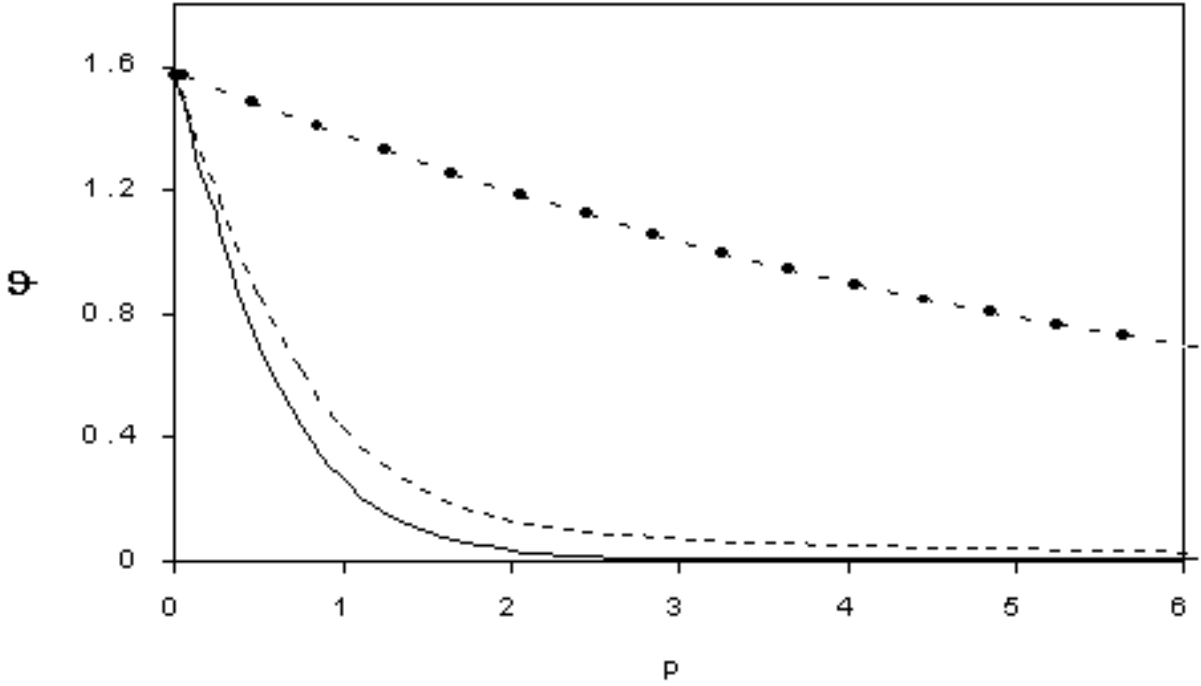}
\caption{General solution of the SD equation (\ref{eq:formla5}) for
potential (\ref{potential}) in dimensionless units and $\alpha_s=0.2$. 
Solid line -- for current quark mass $m_{o}= 0.021 \Vo$;
Dashed line - for $m_{o}= 0.2 \Vo$; dashed point line -- for $m_{o}= 5 \Vo$. }
\end{figure}
\newpage
\begin{figure}
\psfig{figure=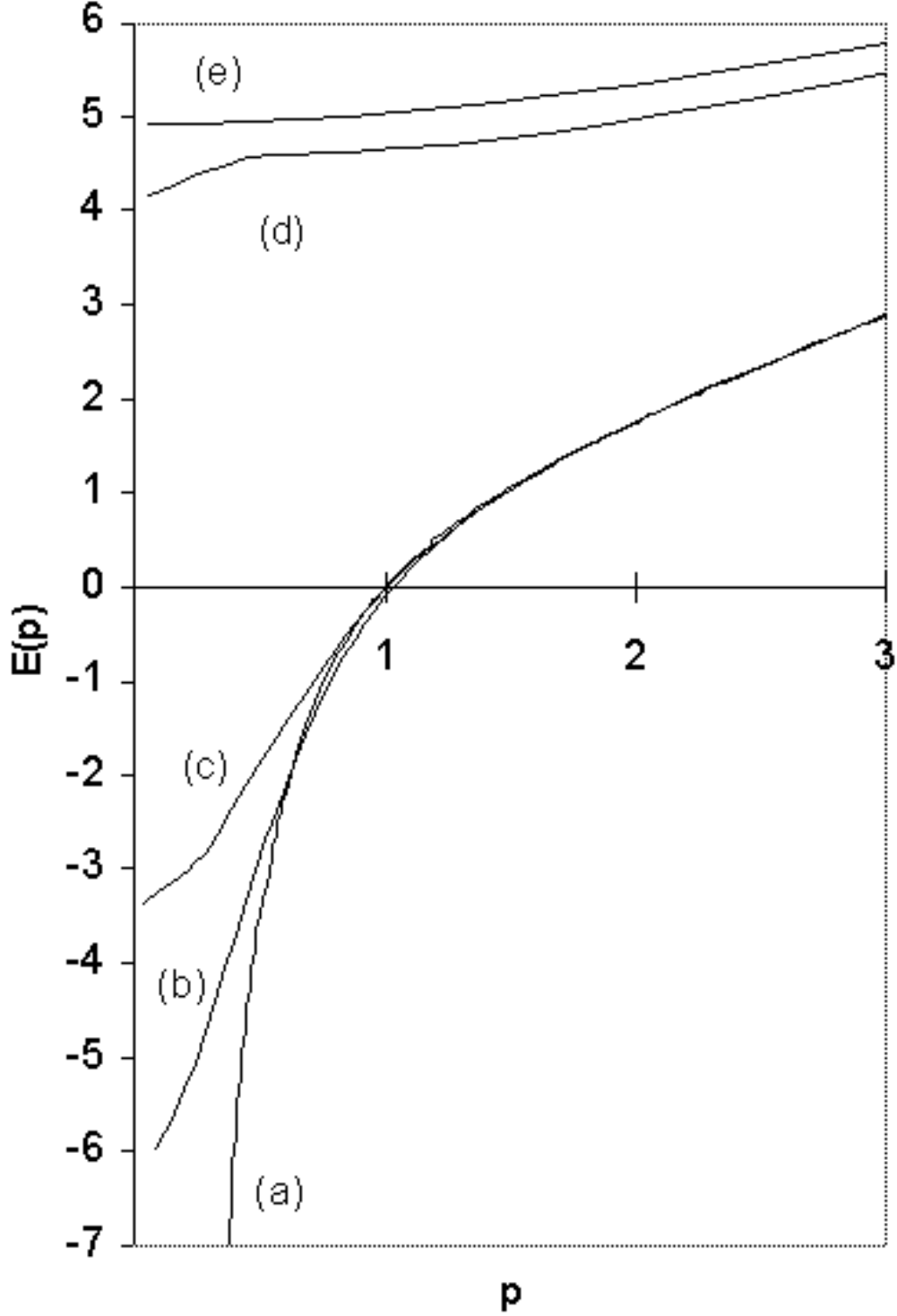}
\caption{Quark self-energy $E(p)$ is calculated from $\varphi(p)$ under
different quark masses in dimensionless unit. 
(a)In the neigborhood p=0 and $\alpha_s=0$ behaviour $E(p)$ from
expression (\ref{eq:formla6}) is ${\bar E}(p) = p -\frac{1}{p^2}$; 
 (b) under $m_o=0, \alpha_s=0$ repeat results Orsay group; 
 (c)massive quark $m_o=0.2 \Vo$, without Coulomb term  potential 
$\alpha_s= 0$; (d) massive quark $m_o=5 \Vo,$ potential as sum oscillator
and Coulomb term $\alpha_s=0.2$; 
 (e) massive quark  $m_o=5 \Vo,$ without Coulomb term $\alpha_s=0$}
\end{figure}
\newpage
\begin{figure} \psfig{figure=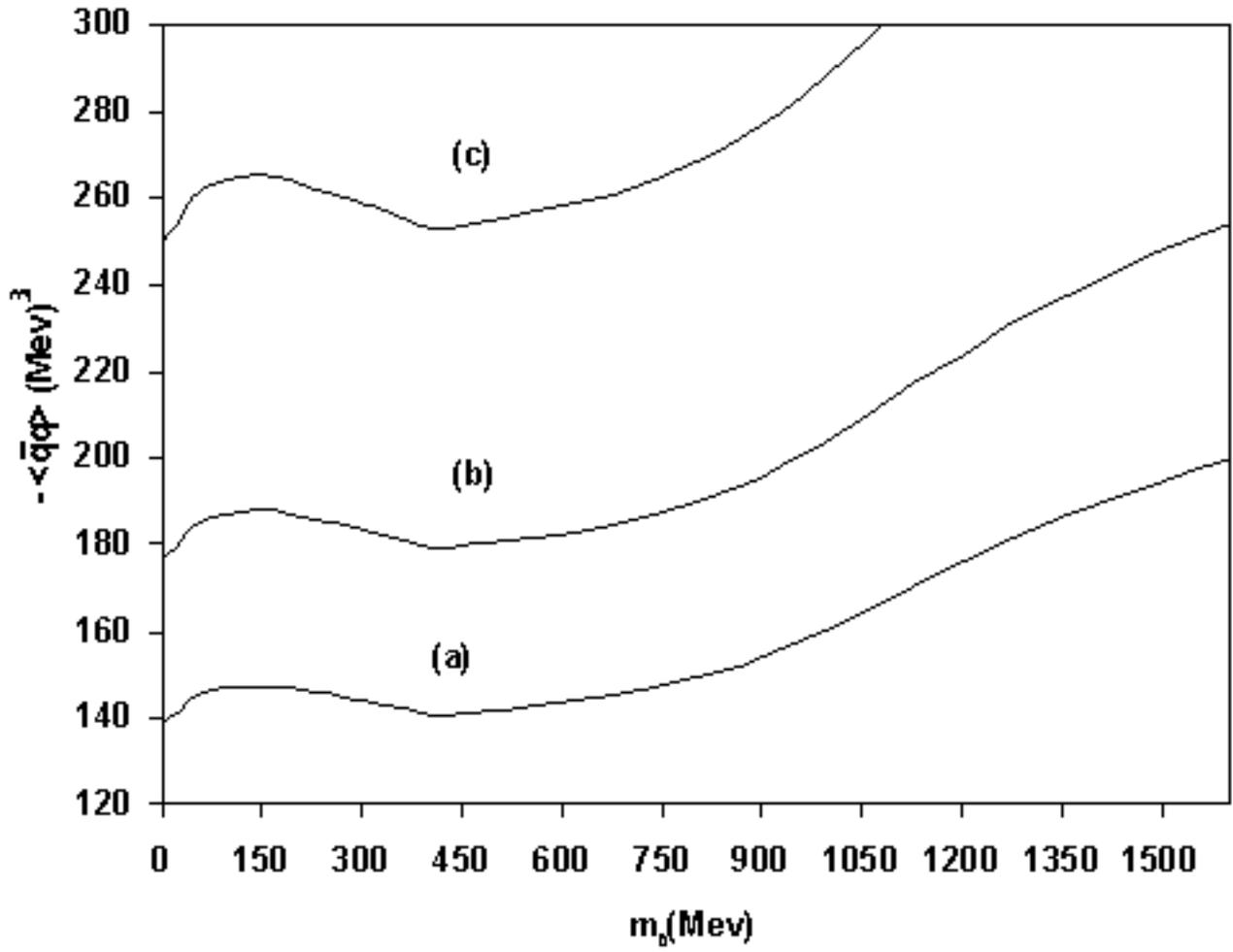}
\caption{The quark condensate under different masses quark $m = m_{o1} +
m_{o2}$ 
(a)under $\Vo=289$ Mev in point $m_o=0$ this results complies with
Orsay group; 
(b) under $\Vo = 368$Mev; 
(c) under $\Vo=520$ Mev.}
\end{figure}
\end{document}